\documentstyle[aps, prb]{revtex}

\def\be{\begin{equation}}
\def\fe{\end{equation}}

\def\Ext{{\text{ext}}}
\def\Int{{\text{int}}}
\def\Mag{{\text{mag}}}
\def\Kin{{\text{kin}}}
\def\Dyn{{\text{dyn}}}
\def\Tube{{\text{tube}}}
\def\Con{{\text{con}}}
\def\ns{n_s}
\def\ninf{n_\infty}
\def\Oinf{\Omega_\infty}
\def\Binf{B_\infty}
\def\Econ{\E_\Con}
\def\Etube{\E_\Tube}

\def\r{r}
\def\V{{\cal V}}
\def\A{{\cal A}}
\def\B{{\cal B}}
\def\E{{\cal E}}
\def\K{{\cal K}}
\def\Bo{{\B_0}}
\def\Emag{\E_\Mag}
\def\Ekin{\E_\Kin}
\def\Umag{U_\Mag}
\def\Ukin{U_\Kin}
\def\Uhat{\widehat U}
\def\xcheck{\check x}
\def\Kcheck{\check K}
\def\echeck{\check \epsilon}

\begin{document}
\draft
\wideabs{
\title{Energy of Magnetic Vortices in Rotating Superconductor}
\author {Brandon Carter, Reinhard Prix, David Langlois}
\address{D\'epartement d'Astrophysique Relativiste et de Cosmologie,\\
Centre National de la Recherche Scientifique, \\Observatoire de
Paris, 92195 Meudon, France.}
\date{August 9, 1999}
\maketitle 

\begin{abstract}
{\bf Abstract.} 
We carry out a systematic analytic investigation of stationary and
cylindrically symmetric vortex configurations for simple models
representing an incompressible non-relativistic superconductor in a
background, which is rigidly rotating with the angular velocity
$\Oinf$. It is shown that although the magnetic and kinetic 
contributions to the energy per unit length of such a vortex are
separately modified by the background angular velocity, its effect on the
total energy per unit length cancels out. For a type II superconductor
threaded by a parallel array of such vortices, this result implies that
the relevant macroscopic magnetic field strength $H$ will not be equal to
the large scale average $\langle B\rangle$ of the local magnetic induction
$B$ (as has previously been suggested) but instead that $H$ will simply be
equal to the external London field value $\Binf = -(2m/q)\Oinf$ (where 
$m$ and $q$ are the mass and charge of the condensate particles)
that characterizes the value of $B$ outside the vortices.
\end{abstract}
}

\section{Introduction}
\label{secIntro}

Because of the property of irrotationality of superfluidity, it is well 
known that the angular momentum of a zero temperature 
superfluid is ``carried'' only by the vortices it contains. Similarly,
when a magnetic field is applied to a type II superconductor, there
exist quantized flux tubes. When one considers a rotating
superconductor, the two effects, velocity and electromagnetism, are
combined. It is the purpose of this work to carefully analyze the
energy of such a system. 

The physical 
motivation behind this work is the need to clear up some confusion that has
arisen in the context of neutron star matter \cite{Men91} about the relation
between the large scale magnetic field strength $H$ and the average value
$\langle B\rangle$ of the local magnetic induction $B$ in a rotating type II
superconductor threaded by a parallel array of vortices. In order
to clarify the issue we shall proceed on the basis of the same kind of
simplification that was postulated as the basis of the earlier
discussion,\cite{Men91} working in terms of a broad category of
non-relativistic incompressible superfluid models that includes, but
is not restricted to, the special case characterized by the standard
Ginzburg-Landau ansatz. 
The main conclusion of our work, as will be shown in Section
\ref{secAverage}, is that the conventionally defined macroscopic field
strength $H$ will simply be given by the value (\ref{12b}) of 
the external (London) limit $\Binf$ of the local induction $B$,
and not by the average value $\langle B\rangle$ as was previously
suggested.\cite{Men91} 

The essential feature of the models to be dealt with is just the usual
postulate that the relevant charged superfluid constituent is represented by
a locally variable number density $\ns$ of bosonic particles that are
characterized by an effective mass $m$, charge $q$ and a momentum
covector having space components 
\be m v_i+q A_i=\hbar\nabla_{\! i}\varphi\, ,\label{0}\fe 
where $\hbar$ is the Dirac-Planck constant, $A_i$ is the
magnetic vector potential and $\varphi$ is a scalar with period $2\pi$
representing the phase variable of the boson condensate. 
In ordinary laboratory applications the particles would represent
Cooper type electron pairs, characterized in terms of the charge and
mass of the electron by the exact relation $q=-2e$, and to a good
approximation by $m\simeq 2m_e$, whereas in the context of
neutron star matter they would represent 
proton pairs, characterized by  $q=2e$ and an effective mass given
roughly in terms of that of the proton by  $m\approx 2m_p$.

The scenarios we shall consider will be of the usual kind, in which each
individual vortex is treated  as  a stationary cylindrically symmetric
configuration consisting of a rigidly rotating background medium with
uniform angular velocity $\Oinf$, say, together with a charged
superfluid constituent  in a state of differential rotation with a velocity
$v$, which tends at large distance towards the rigid rotation value given by
$\Oinf\r$, where $\r$ is the cylindrical radial distance from the
axis . It will be supposed that the superfluid particle 
number density $\ns$ vanishes on the axis and is a monotonically
increasing function of the cylindrical radius variable $\r$, tending rapidly
to a constant value $\ninf$ at large distances from the axis: $\ns=\ninf$
for  $\r\gtrsim \xi$, say, where $\xi$ is a parameter interpretable as the 
core radius.
It will further be supposed that the local charge
density is canceled by the background so that there is no electric field,
but that there is a magnetic induction field with magnitude $B$ and
direction parallel to the axis, whose source is the axially oriented
electromagnetic current whose magnitude $j$ will be given by \be
j=q\ns(v-\Oinf\r) \, .\label{5}\fe 
The relevant Maxwellian source equation for the magnetic field
will have the familiar form
\be {dB\over d\r}= -4\pi j\, .\label{4}\fe
The other relevant Maxwellian equation is the one governing
the axial component $A$
(which in an appropriate gauge will be the only one) of the 
electromagnetic potential covector, which will be related to the
magnetic induction by
\be {d(\r A)\over d\r}=\r B\, .\label{3}\fe
The essential property distinguishing the ``superconducting case'' from
its  ``normal'' analogue is the London flux quantization condition,
which in the present context (where all physically relevant quantities 
depend only on the cylindrical radius $\r$)
will be expressible in the well known form~\cite{Tilley}
\be mv+qA={N\hbar\over\r}\, ,\label{2}\fe
where $N$ is the relevant  phase winding number, which must be an integer.

It is to be noted of course that by themselves the foregoing equations 
are not quite sufficient to fully characterize the
model: in order to obtain a complete system it is also necessary 
to have some well defined prescription for 
the radial dependence of the number density $\ns$, which will be referred
to below as the {\it structure function}.
The available literature does not seem to provide any
fully adequate general purpose ansatz for such a structure function, 
though various, more or less satisfactory, phenomenological
prescriptions have been put forward in  particular contexts. One of
the simplest proposals is   to postulate that $\ns$ falls
discontinuously from its asymptotic constant value $\ninf$ to
zero. Such a simple ansatz is in fact perfectly adequate for many
purposes, since, as will be seen below, much of the relevant physics
turns out to be insensitive to the detailed structure of the core.
However, no such specific prescription for the structure function will
be needed to obtain the  general result of section~\ref{secAverage}.

The plan of this paper is the following. In section \ref{secHomogene},
we transform our system of equations to a simpler form by considering
the deviations of all quantities with respect to their asymptotic
values corresponding to rigid rotation. Section \ref{secLemma} is
devoted to the demonstration of the cancellation between the
rotation--induced terms of the kinetic energy and the magnetic
energy. In section \ref{secAxisfield}, we show that there is a simple
relation between the total energy  per unit length of the vortex and
the total flux independently of the  details of the structure of the
vortex. Section \ref{secExternal} and \ref{secInternal} are concerned 
respectively with the external and the internal solution representing
the vortex. In section \ref{secExplicit} we evaluate explicitly the energy
contributions as functions of the core parameters and finally, in
section \ref{secAverage} we apply our results  for  one vortex to
the case of an array of aligned vortices and obtain our main result
concerning the macroscopic field strength $H$. 
Section \ref{secConclusions} summarizes this work.
 
\section{Homogenization of the system}
\label{secHomogene}
 For given values of the relevant physical constants $m$ and $q$
and the rotation rate $\Omega_{\infty}$, and subject to the provision
that the structure
function for $\ns$ has been prescribed in advance, the foregoing equations will
constitute a linear differential system relating the variable
functions $v$, $B$, $A$ to the integer valued parameter $N$.
Before proceeding, it will be useful to take advantage of the
possibility of transforming the preceding
system of equations to a form that is not just linear but also homogeneous
by replacing the variables $v$, $B$, $A$ by corresponding variables
$\V$, $\B$, $\A$ that are defined by
\be \V=v-\Oinf\r\, , \label{11} \fe
\be \B= B-\Binf\, ,\label{12}\fe
\be \A=A- {_1\over^2}\r \Binf\, .\label{12a}\fe
Here $\Binf$ is the uniform background magnetic field
value that would be generated by a  rigidly  rotating superconductor and is
 given by the London formula,
\be \Binf= -{2m\over q}\Oinf \, , \label{12b}
\fe
obtained by combining (\ref{2}) and (\ref{3}) in the specialized case of rigid
corotation, which is $\V=0$. 

In terms of these new variables the equation (\ref{3}) will be 
transformed to the form
\be {d(\r \A)\over d\r}=\r \B\, ,\label{3c}\fe
while the other differential equation (\ref{4}) will be
transformed to the form
\be {d\B\over d\r}=-4\pi j\, ,\label{13}\fe
in which we shall have
\be j= q\ns\V\, .\label{2b}\fe
Finally the flux quantization condition (\ref{2}) will be converted to the 
form
\be m\V+q\A={N\hbar\over\r}\, ,\label{2c}\fe
which can be used to transform (\ref{3c}) to the form
\be 
{m\over q\r}{d (\V\r)\over d\r}=-\B\, .\label{14}
\fe
The advantage of this reformulation is that unlike $v$, $B$ and $A$,
the new variables $\V$, $\B$ and $\A$ are subject just to
homogeneous boundary conditions: they must all
tend to zero as $\r\rightarrow\infty$, while at the inner boundary,
as $\r\rightarrow 0$, there is just the regularity
requirement that $\B$ should be bounded, so that we have
$\B\rightarrow \Bo$ for some finite limit
value $\Bo$, which by (\ref{3c}) entails automatically
that $\A$ should tend to zero.
Since the number density $\ns$ is postulated to vanish
at the origin there is no corresponding restriction on $\V$.
We have thus obtained a homogeneous linear system of equations
relating the integer $N$ to the set of three
functions consisting of the excess (with respect to the background)
magnetic induction variable $\B$, and the corresponding excess
potential variable $\A$ together with the relative velocity variable
$\V$, or equivalently the current magnitude $j$ as given by (\ref{2b}).
This means that they will be expressible in the form
\be \B=N\tilde\B\, , \hskip 1 cm \A=N\tilde\A\, ,
\label{19a}\fe
\be \V=N\tilde V\, ,\hskip 1 cm j=N\tilde j\, ,\label{19b}\fe
in terms of corresponding
rescaled functions $\tilde\B$, $\tilde\A$,
$\tilde\V$ and $\tilde j$ that will be fully determined 
(independently not just of the rotation parameter $\Binf=-2m
\Oinf/q$ but also of the winding number $N$) just by the
physical constants $m$ and $q$ and the specification of the
structure function giving the radial dependence of the number 
density $\ns$.

\section{Rotation energy cancellation lemma}
\label{secLemma}
One of the main purposes of the present work is to demonstrate,
in the present section, a useful lemma concerning 
mutual cancellation -- {\it independently of the radial dependence}
of the relevant particle density $\ns$ --
between the background rotation
 dependent term in the magnetic energy per unit
length 
\be \Umag =\int \Emag\, 
dS\, ,\label{E1}\fe 
and the corresponding  term in  the kinetic energy per unit length
\be \Ukin =\int\Ekin\, 
dS\, ,\label{E2}\fe 
with
\be dS=2\pi\r\, d\r\, .\fe
In the above expressions,
 $\Emag$ is the extra magnetic energy density arising
from a non-zero value of the phase winding number $N$, i.e., the local
deviation from the magnetic energy density due just to the uniform field
$\Binf$ associated with  the state of rigid corotation at the
 angular velocity $\Oinf$, namely
\be \Emag=
{B^2\over 8\pi}-{\Binf^{\, 2}\over 8\pi}
\, ,\label{32}\fe
while $\Ekin$ is the corresponding deviation of the kinetic
energy from that of the state of rigid corotation at the
 angular velocity $\Oinf$, namely
\be \Ekin=
{m\over 2}\ns\big(v^2-\Oinf^{\, 2}\r^2\big)
\, . \label{33}\fe
Note that in addition to $\Umag$ and $\Ukin$ the total energy per unit
length $U_\Tube$ associated with the vortex will contain an extra
potential energy term allowing for effect of the breakdown of superfluid
condensation in the core, but this will not be relevant for the work of the
present section. In the limiting case of an ordinary superfluid, as
characterized by vanishing charge $q=0$, the kinetic contribution would be the
dominant one, but in the context of superconductivity, i.e., when $q$ is
non-zero, it is commonly~\cite{Men91} overlooked, perhaps because of the small
value of the electron mass that is relevant in laboratory applications. The
purpose of the present section is to show not only that the kinetic
contribution will not in general be negligible compared with the magnetic
contribution, but also that its inclusion  brings about
considerable simplification.

To start with, using the the decomposition (\ref{12})  of the magnetic
field, it will be possible to express the magnetic energy density
contribution in the form
\be \Emag=
{\B^2\over 8\pi}+{\Binf\B\over 4\pi}
\, ,\label{32a}\fe
while similarly, using the decomposition (\ref{11}) of the velocity,
it will be possible to express the corresponding kinetic
energy density in the analogous form
\be \Ekin=
{m\over 2}\ns\big(\V^2+2\Oinf\V\r\big)
\, ,  \label{33a}\fe
which can usefully be rewritten in terms of the current magnitude $j$, 
using (\ref{2b}) and (\ref{12b}) as
\be \Ekin=
{m\over 2}\ns\V^2-{j\over 2}\Binf\r
\, .  \label{33b}\fe
Using the Maxwell source equation (\ref{13}) this can be converted to the
form
\be \Ekin= {m\over 2}\ns\V^2+{\Binf\over\r}{d\over d\r}\Big({\r^2\B\over8\pi}\Big) 
-{\Binf\B\over 4\pi}\, ,\label{33c}\fe
in which the second term can be seen to be a pure divergence, while
the last term can be seen to be equal in magnitude but opposite
in sign to the last term in (\ref{32a}), so that there will be a 
cancellation between them when the magnetic and kinetic contributions
are combined.

At an integrated level, in view of (\ref{32a}), 
it will  be possible to express the
magnetic energy in terms of quantities 
$\Uhat_\Mag$ and $\Phi$ that are specified
independently 
of $\Binf$, in the form
\be \Umag=\Uhat_\Mag+{\Binf\over 4\pi}
\Phi\, ,\label{37}\fe
where the part that would still be present if the background were 
non rotating is given by
\be \Uhat_\Mag=\int {\B^2\over 8\pi}\, dS\, ,
\label{38}\fe
and where the coefficient $\Phi$ is a flux integral
of the simple form
\be \Phi=\int \B\, dS=N\phi\, ,\label{39}\fe
where $\phi$ is the usual flux quantum, given by
\be\phi={2\pi\hbar\over q}
\, .\label{21}\fe
In a similar manner, it will be possible to express the kinetic contribution 
in terms of quantities $\Uhat_\Kin$
and $\Phi_\Kin$ that are also specified independently of 
$\Oinf$, or equivalently of $\Binf$,
 in the form
\be \Ukin= \Uhat_\Kin+{\Binf\over 4\pi}
\Phi_\Kin\, ,\label{42}\fe
where the part that would still be present if the background were 
non rotating is given by
\be \Uhat_\Kin={m\over 2q}\int j \V\, dS\, ,
\label{43}\fe
and where the coefficient $\Phi_\Kin$ is 
given by
\be \Phi_\Kin=\pi\int \r^2 {d\B\over d\r}\, d\r
=\pi\int d\big(\r^2\B\big)-\int
2\pi\r\B\, d\r\, ,\label{45}\fe
which corresponds to  the last two terms on the right hand side of (\ref{33c}).
The first term in this  expression clearly vanishes when the integration
is taken over the whole range from the center, where $\r=0$, to the
large radius limit where $\r^2\B\rightarrow 0$, as can be seen from the 
explicit solution (\ref{19}). 
We are thus left with the second term, the kinetic analogue
of the magnetic flux contribution,
to which it is evidently equal in magnitude but opposite in sign,
i.e., we obtain
\be \Phi_\Kin=-\Phi\, .\label{46}\fe
It can thus be seen that there is a remarkable cancellation whereby
the dependence on $\Binf$, or equivalently on $\Oinf$,
in the separate magnetic and kinetic energy contributions will
cancel out when they are combined, so that we are left
simply with a result of the form
\be  \Umag+\Ukin
=\Uhat_\Mag+ \Uhat_\Kin
=\int\Big( {\B^2\over 8\pi}+{m\over 2}\ns
\V^2\Big)\, dS 
\, .\label{47}\fe
Since the terms in this expression are both  quadratically dependent
on fields, namely  $\V$ and $\B$, that by (\ref{19a})
and (\ref{19b}) will just be proportional to the winding number $N$, we
obtain the following conclusion.

{\bf Rotation energy cancellation lemma:}
Whereas the separate values of the the magnetic and kinetic contributions (as
defined using the formulae (\ref{37}) and (\ref{42}) above) to the energy per
unit length of the vortex will be affected by  the rate of rotation of the
background $\Oinf$ (or equivalently the corresponding 
London field $\Binf=-2m\Oinf/q$), the combination of these two
contributions {\it will not depend directly on $\Oinf$} 
and can be simply expressed in the form (as a result of equations (\ref{19a})
and (\ref{19b}))
\be  \Umag+\Ukin =\tilde U N^2\, ,\label{47b}\fe
where we recall that $N$ is the winding
number and $\tilde U$ depends  only on the 
physical constants $m$ and $q$ and on the form
of the radial distribution of the number density $\ns$. 
A simple form for $\tilde U$ will be given in the next section.

\section{Axis-field energy formula}
\label{secAxisfield}
The preceding result, namely the cancellation of the contributions due 
to the  background rotation, was obtained simply with the 
background London equation (\ref{12b})  without using the full, i.e., local 
London quantization condition
(\ref{2}). The ultimate cancellation of the $\Binf$~--~dependent 
contribution is attributable at a local level  to the fact
that the $\Binf$~--~dependent contribution to the 
combined energy density is a
pure divergence: 
\be 8\pi\big(\Emag+\Ekin\big)
= \B^2+4\pi{m\over q}j\V+{\Binf\over\r}
{d\over d\r}\big(\r^2\B\big)\, .\label{47c}\fe
We can obtain a stronger result if we now invoke the more specialized relation
(\ref{14}) which is a consequence of the quantization condition
(\ref{2}) that specifically characterizes superconductivity.
This condition can be seen to imply that the whole of the right 
hand side of (\ref{47c}) will be expressible as a divergence, since 
we shall have 
\be \B^2+4\pi{m\over q}j\V=-{m\over q\r}
{d\over d\r}\big(\r\V\B\big)\, .\label{49}\fe
It can thereby be seen, using (\ref{2}) again, 
that the combined energy density will be expressible as
\be \Emag+\Ekin
={1\over8\pi\r}{d\over d\r}\left(\Big(\Binf\r^2
+\r A-{N\hbar\over q}\Big)\B\right) \, .\label{50}\fe
In the outer limit, as $\r\rightarrow\infty$, the rapid fall off of
$\B$ will ensure that the quantity inside the divergence will tend
to zero. In the inner limit, as $\r\rightarrow 0$, the
first term in the divergence obviously gives no contribution, and the 
consideration that $A$ should be bounded ensures that the second term also
gives no contribution, so we shall be left with the contribution
just from the final term, which is proportional to the winding number
$N$. The final outcome of the integration of (\ref{50})
can be stated as follows.

{\bf Axis-field energy lemma:}
Subject to the London quantization (as given by (\ref{2}) above) the
combination of the magnetic and kinetic contributions (as defined using the
formulae (\ref{37}) and (\ref{42}) above) to the energy per unit length 
for a vortex with given winding number $N$ and corresponding total
flux $\Phi$ as specified by (\ref{39}) will be provided just by the 
the axis-field value $\Bo$ according to the proportionality law 
\be \Umag + \Ukin
= {\Phi\Bo\over 8\pi}\, , \label{51}\fe
where $\Bo$ is the  value on the axis of 
the {\it relative} magnetic field 
value $\B$ as given by (\ref{12}), i.e. it is the difference 
\be \Bo=B_0-\Binf\, \label{52}\fe
between the central value, $B_0$, of the
magnetic induction $B$ and its asymptotic London value $\Binf$. 

A corollary of this  second lemma is  
that the combination of the  kinetic and magnetic energy per unit length
 will remain the same whatever the internal structure, as long as 
the total flux and the axis magnetic field are the same. The simplest 
such configuration is given by  a field $B$
retaining the same uniform central value $B_0$ out to a cut--off where it
drops discontinuously to its asymptotic value $\Binf$. This cut--off
radius, $\tilde R$, say, is adjusted so as to give the same total  
flux as in the actual model, i.e. so as to satisfy the specification
\be \pi \tilde R^2={\Phi\over \Bo}=
{\phi\over\tilde\Bo}\, .\label{53}\fe
Since the quantity $\tilde\Bo$ , i.e. the value on the
axis of the rescaled field defined by (\ref{19a}), depends only
on the physical constants $m$ and $q$ and on the form of the
structure function
specifying the radial dependence of the number density $\ns$,
it follows that the same applies to the effective
 radius $\tilde R$, which will thus be 
independent of $N$, as well as of
the background rotation rate $\Oinf=-q\Binf/2m$.
The conclusion that the effective magnetic radius depends
only on the structure function specifying $\ns$ is interpretable
as a restatement of our first lemma, since it can be seen 
that the coefficient $\tilde U$ in (\ref{47b}) will be given
just in terms of this effective radius $\tilde R$ by the formula
\be \tilde U={\hbar^2\over 2 q^2 \tilde R^{\,2}}\, .\fe

\section{Average over an array of aligned vortices}
\label{secAverage}
\medskip

Let us now consider the typical situation in a type II superconductor,
 in which we have not just a single vortex
but a parallel array of such vortices with
sufficiently low mean number density per unit surface area,
$\nu$, say, for the separation distance 
between neighboring vortices to be large compared with the penetration length
$\lambda$. Since, according to (\ref{12}) and (\ref{39}), each vortex carries an extra
magnetic flux $\Phi$ in addition to the contribution from
the uniform London field $\Binf$, the large scale average magnetic field
will be given by
\be \langle B\rangle=\Binf+\nu\,\Phi
\, .\label{92}\fe
 As compared with the average energy density of a configuration in
rigid corotation with the given angular velocity $\Omega_\infty$, but with
no magnetic field,
the extra energy density averaged over a large number of vortices
will be given by
\be \langle\E\rangle= \E_{\text{Lon}}+ \langle\Etube\rangle
\, ,\label{93}\fe
where $\E_{\text{Lon}}$ is the uniform  contribution
from the London magnetic field, i.e.
\be  \E_{\text{Lon}}={\Binf^2\over 8\pi}\, ,\label{94}\fe 
and where
$\langle\Etube\rangle$ is the large scale average of the
contribution given locally for the separate vortices by 
\be \Etube= \Emag+\Ekin
+\Econ\, .\label{95}\fe
where $\Emag$ and $\Ekin$ are the magnetic
and kinetic energy contributions discussed in the preceding sections
and $\Econ$ is the condensation energy contribution
depending 
just on the radial distribution of the condensate number density
$\ns$ (in a manner that is unimportant for our
present  purpose), which means that the corresponding additional contribution 
\be U_{\text{con}} =\int \Econ\, 
dS \label{E3}\fe 
to the vortex energy per unit length can be treated just as a constant
as far as the present section is concerned. It follows that we shall have
\be\langle\Etube\rangle= \nu U\, ,\label{96}\fe
where $U$ is the total energy per unit length of an individual vortex as 
given by the combination
\be U=\Umag+\Ukin+U_\Con\, ,\label{97}\fe
in which the first two contributions will separately
depend on the background
rotation velocity $\Oinf$ (or equivalently on the London
field $\Binf$) but in which, by the
cancellation lemma expressed by (\ref{47}), the total (like 
final term) will not.

Since each vortex is associated
with a momentum circulation of magnitude $2\pi\hbar N$ there will
be a corresponding generalized vorticity, in the sense of momentum
circulation per unit area, with large scale average given by
\be \langle w\rangle=2\pi\hbar N\nu\, .\label{98}\fe
In terms of this quantity the large scale average (\ref{93}) 
of the extra energy due
to the deviation from a configuration of unmagnetized rigid
 corotation will be given by
\be \langle\E\rangle= {\Binf^{\,2}\over 8\pi}+{\langle w\rangle
\over 2\pi\hbar}{U\over N}\, .\label{99}\fe
For a large scale variational description it is convenient to use
$\langle w\rangle$ and $\langle B\rangle$ as the independent variables.
In terms of these, the London field can be seen from (\ref{92}) 
to be expressible as
\be 
\Binf=\langle B\rangle-{\langle w\rangle\over q}\, ,\label{100}
\fe
so (\ref{99}) gives
\be \langle\E\rangle= {\langle B\rangle^2\over 8\pi}
-{\langle B\rangle\langle w\rangle\over 4\pi q}+{\langle w\rangle^2\over
8\pi q^2} +{\langle w\rangle\over 2\pi\hbar} {U\over N}\, .\label{101}\fe
Since, by (\ref{51})  the ratio $U/ N$ will be given by the formula
\be {U\over N}
= {\tilde\Phi\Bo\over 8\pi}+{U_{_{\rm con}}\over N} 
\, ,\label{102}\fe
in which all dependence on $\Binf$ and thus also on $\langle B\rangle$
has canceled out, it can immediately be seen that the conventional definition
(which is the same as  the definition adopted in Ref.~\onlinecite{Men91})
\be H =4\pi {\partial \langle \E\rangle\over
\partial \langle B\rangle}\, \label{103}\fe
for the effective magnetic field strength $H$ will simply give
\be H=\Binf\, ,\label{104}\fe
i.e. $H$ is directly identifiable with the London field.
The corresponding magnetic polarization ${\cal M}$, as defined in
the usual way by
\be \langle B\rangle=H+4\pi {\cal M}\, \label{105}\fe
will be expressible as
\be {\cal M}={\langle w\rangle\over 4\pi q}={\nu N\hbar\over 2q}
\, .\label{106}\fe

\section{External solution}
\label{secExternal}
In the previous sections we have been able to establish very useful properties
concerning the energy density per unit length of a vortex, without 
needing the specification of a internal structure. This section and the next 
one will consider this question and show in particular how the unspecified 
parameter of the previous section, the axis value of the relative magnetic 
field, or equivalently the effective radius $\tilde R$, can be explicitly
computed depending on the modelization of the internal structure. 
We shall focus here on the solution outside the core, which is always the 
same up to a normalization constant, that can be determined only with the
knowledge of the internal structure. This will be the purpose of the next 
section.

In the region outside
the core, i.e. in the range $\r\geq \xi$, where the the number
density $\ns$ is uniform with value
\be \ns=\ninf\, ,\fe
the equation obtained from (\ref{13}) and (\ref{14})
by eliminating $\B$ will have the form
\be \r^2{d^2\V\over d\r^2}+\r{d\V\over d\r}-\left({\r^2\over\lambda^2}
+1\right)\V=0\, ,\label{15}\fe
where  $\lambda$ is a fixed lengthscale given by
\be \lambda^2={m\over 4 \pi q^2 \ninf}\, \label{16}\,,\fe
and which is called the London penetration length.

The equation (\ref{15}) is of the well known Bessel--type, whose most general
asymptotically bounded solution is expressible
 in the form (see Ref.~\onlinecite{as})
\be \V=C K_1\{x\}\, ,\label{17}\fe
where the independent variable $x$ is defined by
\be x={\r\over\lambda}\, ,\fe
$C$ is a normalization constant and $K_1$ is
a modified Bessel function. 

It follows immediately from the flux quantization condition
(\ref{2c}) that the magnetic potential
deviation defined by (\ref{12a}) will be given by 
\be
\A=-{m\over q}CK_1\{x\}+{\Phi\over 2\pi\r}\, ,\label{20}
\fe
where $\Phi$ is the magnetic flux integral given by (\ref{39}).

Using the fact that $K_1$
 is related to the Bessel function $K_0$ by
\be K_1=-K_0^\prime\, , \hskip 1 cm 
K_0=-K_1^\prime-x^{-1}K_1\, ,\label{18}\fe
where a prime stands for differentiating with respect to the 
argument $x$,
it is straightforward  to obtain the corresponding solution of (\ref{14})
for the magnetic field deviation,
which will be expressible in the simple form
\be \B={m\over q\lambda}C K_0\{x\}\, .\label{19}\fe 
The external configuration for the magnetic vortex has thus far been
determined up to the normalization constant $C$. It will be seen in 
the next
section how, on the basis of a suitable ansatz for the radial dependence
of $\ns$,  the solution inside the core can be used to fix this
constant $C$, and thus to determine completely the configuration of the
vortex.

\section{Internal solution}
\label{secInternal}
Instead of directly specifying the way in which the number density $\ns$
varies from zero on the axis ($r=0$) to its external value $\ninf$ at the
core radius (where $\r=\xi$, i.e. where $x=\check x\equiv \xi/\lambda$), 
it is more convenient to work with an ansatz based on an explicit
prescription for the current magnitude $j$, which will have a 
qualitatively similar behavior in the core, ranging from zero on 
the axis to a value $\check j$ at the core radius, 
that according to (\ref{2b}) and (\ref{17}) will be given by
\be \check j=q \ninf C\check K_1\, ,\label{61}\fe
using the obvious abbreviation $\check K_1=K_1\{\check x\}$.
The current  in the core, i.e. where
$x\leq \check x$, will therefore be expressible in the form
\be j=\sigma\check j\, ,\label{62}\fe
where $\sigma$ is a dimensionless function of $x$ that is required to
vanish, $\sigma=0$, for $x=0$ and to increase to unity, $\sigma=1$
where $x=\check x$ ($\sigma$ plays here the role 
of the structure function mentioned in the introduction). 
 For any suitably prescribed function $\sigma$ with
these properties, there will be corresponding functions, $\chi$ and
$\zeta$, say, that are defined by the requirement that they too
should vanish on the axis, i.e., $\chi=\zeta=0$ for $x=0$ and by the
requirement that they should be obtained in the region $0\leq x\leq\check x$ as
solutions of the differential equations
\be \xcheck\, {d\chi\over d x}= 2\sigma\, ,\hskip 1 cm
\xcheck \, {d(x\zeta)\over d x} = 4x\chi\, .\label{63}\fe
In terms of such a set of functions, the relevant solution of equation
(\ref{13}) will evidently be given by
\be \B=\B_0-2\pi \xi\check j\,\chi\, ,\label{64}\fe
and the corresponding solution of (\ref{3c}) will be given by
\be\A={_1\over^2}\lambda \Bo\, x-{_1\over^2}\pi \xi^2
\check j\,\zeta 
\, .\label{65}\fe
The requirement that the magnetic field should be continuous
(so that the current density $j$ remains finite)
entails that the internal solution (\ref{64}) should match
the corresponding external solution (\ref{19}) where $x=\check x$,
so we obtain a boundary condition of the form
\be  \B_0-2\pi \xi\check j\, \check\chi=
{m\over q\lambda}C \check K_0\, ,\label{66}\fe
while the corresponding continuity requirement for the potential gives
a second boundary condition of the form
\be \Bo-\pi \xi\check j\,\check\zeta ={\Phi\over\pi \xi^2}-
{2m\over q\xi}C\check K_1\, .\label{67}\fe
This pair of boundary equations can be solved to give the central
magnetic field difference in the form
\be \Bo={m\over q\lambda}C\big(\check K_0+
{\check\chi\over 2}\check x
\check K_1\big)\, ,\label{68}\fe
while the required normalization constant $C$ is finally obtained in the 
form
\be C= {N\hbar\over\lambda m \check\K} \, ,\label{69}\fe
in terms of a dimensionless quantity that is given by 
\be \check\K=\big(1+{\check\epsilon\over 8}\check x^2\big)
\check x \check K_1+{_1\over^2}\check x^2 \check K_0=
{\xcheck^3 \over 8}\left(\echeck\,\Kcheck_1 + {4\over\xcheck}\,\Kcheck_2\right)
\, ,\label{27}\fe
in which the only dependence on the internal structure is that
embodied in the dimensionless number $\check\epsilon$ which can be
seen to be given in terms of the boundary values $\check\chi$ and
$\check\zeta$ of the functions $\chi$ and $\zeta$ by the
simple formula
\be \check\epsilon=2\check\chi-\check\zeta\, .\label{70}\fe
This quantity $\check\epsilon$ can be interpreted as the value
at the core boundary $x=\check x$ of a function $\epsilon$ of $x$
given by
\be\epsilon= 2{\check\chi\over\check x} x-\zeta\, ,\fe
in terms of which the solution for the potential difference $\A$
will be given by
\be \A={m C\over 2 q}\big(\check K_0\, x +{\check x^2\over 4}
\check K_1\,\epsilon\big) \, .\fe
The corresponding expression for the magnetic field excess
will have the form
\be \B={mC\over q\lambda}\big(\check K_0+{\check x\over 2}
\check K_1(\check\chi-\chi)\big)\, .\fe
Using the solution (\ref{69}) for $C$, the central value needed
 for the energy formula (\ref{51}) can be seen to be obtainable as
\be \Bo={N\hbar\over q\lambda^2\check x^2}\Big(2-
{\check x\check K_1(8-\check\zeta\check x^2)\over 4\check\K}\Big)
\, .\label{axisfield}\fe

\section{Explicit energy contributions}
\label{secExplicit}

Using the axis--field energy formula (\ref{51}) together with the
solution (\ref{axisfield}) for the axis--field $\Bo$, we can
immediately obtain the total ``dynamical'' energy  $U_\Dyn \equiv
\Umag + \Ukin = \Uhat_\Mag + \Uhat_\Kin$  (which is the total energy
less the condensation energy): 
\be
U_\Dyn = U_0\,
\frac{2\Kcheck_0 + \xcheck\check\chi\Kcheck_1}
{2\check\K},
\label{Udyn}
\fe
where
\be  
U_0\equiv\left({N\phi \over
4\pi\lambda}\right)^2.
\fe 
Using the external solution for $\B$ and $\V$, one can  obtain
the external magnetic and kinetic energy contributions, defined in
(\ref{38}) and (\ref{43}), in the form 
\be
\Uhat^\Ext_\Mag = \left({mC\over 2q}\right)^2\,{\xcheck^2\over 2}\,
\left({\Kcheck_1}^2 - {\Kcheck_0}^2\right) 
\fe
and
\be
\Uhat^\Ext_\Kin = \left({mC\over 2q}\right)^2\,{\xcheck^2\over 2}\,
\left[\,{2\over\xcheck}\,\Kcheck_0\,\Kcheck_1 - 
\left({\Kcheck_1}^2 - {\Kcheck_0}^2 \right) \right],
\fe
where the core structure dependence is contained exclusively in the 
constant $C$.
This constant $C$, using the solution (\ref{69}), can be related to the (core structure independent) 
constant $U_0$ defined just above by the simple relation
\be
\left({mC\over 2q}\right)^2={ U_0\over {\check\K}^2},
\fe
where, obviously, all the dependence on the core structure is contained 
in the dimensionless term ${\check\K}$. 

As a consequence,  the total external energy contribution can be rewritten 
in the simple form
\be
\Uhat^\Ext = \frac{\xcheck\,\Kcheck_0\,\Kcheck_1}{{\check\K}^2}\,U_0,
\fe
which can be decomposed in two similar expressions for the magnetic and 
kinetic contributions. 
The  sum of the magnetic and kinetic  {\it internal }contributions can then 
 of course be obtained by using the relation
$\Uhat^\Int = U^\Dyn - \Uhat^\Ext$.

\section{Conclusions}
\label{secConclusions}
\medskip
Let us summarize the results of the present work.
We have first shown that the contributions linearly dependent on 
$\Oinf$ in the magnetic and kinetic energies {\it cancel} each other.
As a consequence, we find that the effective magnetic field strength $H$ 
is simply the London field. 
It is to be observed that the extra energy density contribution arising 
from the second term in (\ref{37}) would give an extra contribution of the form
$\Binf\langle w\rangle /4\pi q$ in (\ref{101}). By including this extra term
-- overlooking the fact that, according to (\ref{46}), it will be 
canceled by the second term in the kinetic contribution (\ref{42}),
which was not taken into account -- the analysis by Mendell
\cite{Men91} provided the erroneous conclusion 
that there would be no polarization, or in other words  
that $H$ should be identified not with the London
field but simply with the mean induction, meaning the
replacement of (\ref{104}) by $H=\langle B\rangle$.

The identification (\ref{104}) of $H$, as given by the conventional
definition (\ref{103}), with the asymptotic London field $\Binf$
has been established here as a precise mathematical relation
in the framework only of a particularly simple model. The problem
of generalization to more sophisticated models, allowing for
compressibility, relativistic effects and other relevant
complications, remains to be dealt with in future work. 

\acknowledgments
We wish to thank R.~Combescot, L.~Lindblom and E.~Varoquaux for
instructive discussions.

\end{document}